\documentclass[aps,prc,preprint,showpacs,groupedaddress]{revtex4-1}
\usepackage{graphicx}
\usepackage{epsfig}

\begin{document}

%\preprint{}

\title{Experimental evidence of the $^6$He level at $E^*$ = 18.3 MeV by the $^4$He+$^3$H three-body reaction}

% author. planatory tt should go in the []'s, actual e-mail

\author{O.M. Povoroznyk}\thanks{orestpov@kinr.kiev.ua} \author{O. K. Gorpinich} \author{O.O.~Jachmenjov}  \author{H.V. Mokhnach} \author{ O. Ponkratenko}
%\thanks{}
\affiliation{Institute for Nuclear Research, National Academy of Science of Ukraine, 03680 Kiev, Ukraine}

\author{G. Mandaglio$^{1,2}$, F. Curciarello$^{1,2}$, V. De Leo$^{1,2}$, G. Fazio$^{1,2}$,} \author{G. Giardina$^{1,2}$}\thanks{ggiardina@unime.it}
\affiliation{$^1$Dipartimento di Fisica dell'Universit\`a di Messina, 98166 Messina, Italy\\ $^2$ Istituto Nazionale di Fisica Nucleare, Sezione di Catania, 95123 Catania, Italy }


\begin{abstract}
Measurements of the t-t and p-t coincidence events in the $^3$H ($\alpha$, ttp) reaction have been obtained at $E_\alpha$ incident energy of 67.2 MeV. Various appropriate angular configurations of detectors were chosen in order to observe the population of the $^6$He$^*$ state at around 18 MeV. Its contribution  appears at the $E_{\rm tt}$ relative energy of 6.0 MeV by the analysis of  bidimensional spectra. We found the formation of the $^6$He excited state at $E^* = 18.3 \pm 0.2$ MeV (with a $\Gamma$ width of 1.1 $\pm$ 0.3 MeV) by the decay into the t+t binary channel, since the threshold energy of the t+t channel is 12.31 MeV. In each analyzed bidimensional energy spectrum of ($E_{\rm t}$,~$E_{\rm t}$) and ($E_{\rm p}$,~$E_{\rm t}$) coincidence events  resonance structures are present due to the formation of both $^6$He$^*$ and $^4$He$^*$ excited states. Our results on the $E^*$ and $\Gamma$ values regarding the $^6$He$^*$ level of  about 18 MeV   are compared with the results obtained by other reactions. Moreover, we also found new $\Gamma$ width values of 0.7 $\pm$ 0.3 and 0.8  $\pm$ 0.4 MeV for the 14.0 $\pm$ 0.4 and 16.1  $\pm$ 0.4 MeV $^6$He levels, respectively.
\end{abstract}


%\keywords{}
\date{\today}
\maketitle

\section{Introduction}
Knowledge of the $E^*$ energy and $\Gamma$ width spectroscopic parameters of excited states of  light nuclei is suitable to test the nuclear models and also to develop astrophysical studies. The best way to measure the above parameters is to study  three-body reactions. In fact, the $E^*$ and $\Gamma$ values deduced by two-body scattering experiments are systematically different from the ones measured when the same states are produced by  three-body reactions as two-body resonances in presence of a spectator particle. Such discrepancies are caused by a relevant background of detected particles due to other concurrent reaction  channels present in the inclusive spectra (single spectra). 

   In a recent paper \cite{pov11}  using a three-body reaction we found the formation of two $^6$Li states at excitation energies around 21 MeV by the decay into two $^3$He+$^3$H  clusters ($\tau$+t) each composed of three nucleons. This result was obtained by investigating the $^3$H($\alpha$,~$^3$H~$^3$He)n  kinematically complete experiment at $E_\alpha$ beam energy sufficient to populate the excitation energy region of our interest. 
   
\begin{figure}[h]
\includegraphics[scale=0.4]{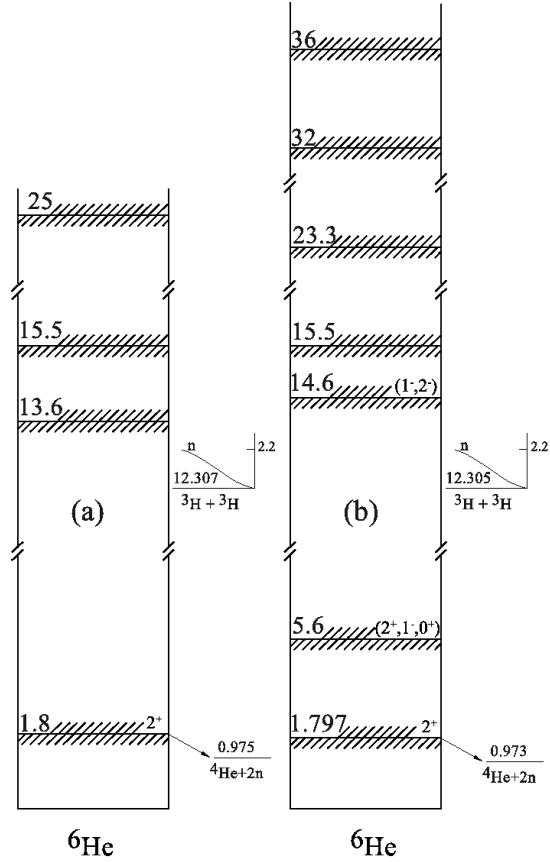} 
\caption{Diagram of the $^6$He energy levels in the Ajzemberg-Selove (a)\cite{selov88}  and Tilley et al. (b)\cite{till02}   compilations.}%\vspace{-0.5cm}
\label{fig1}
\end{figure}

   As Fig.~\ref{fig1} shows, the $^6$He energy level distributions reported in the Ajzemberg-Selove \cite{selov88} and Tilley {\it et al.} \cite{till02}   compilations present some differences, even if we have to underline that compilation \cite{selov88} appeared in 1988, compilation \cite{till02} appeared in 2002 and nowadays the results obtained by other investigated reactions have enriched the set of  possible comparisons. In the diagram of $^6$He levels of compilation \cite{selov88} one low-lying  state appears while  in  compilation \cite{till02}  two states appear which can decay into the $\alpha$+2n channel; moreover, above the  threshold energy of 12.31 MeV for the $^6$He$^*$ states which can decay into t+t clusters, compilation \cite{selov88} gives only three levels up to 25 MeV of excitation energy while compilation \cite{till02} gives five levels up to 36 MeV.

Moreover, in the study of the $^6$Li($^7$Li, $^7$Be)$^6$He reaction \cite{Yamagata05} at incident energy of 455 MeV the $^6$He$^*$  state at  $E^* = 18.0\pm 1.0$ MeV  has been observed with a $\Gamma$ width of $9.5 \pm 1.0$ MeV by the $^6$He$^*$ decay  into the t+t channel, while this $^6$He excited state is not present in either \cite{selov88,till02} compilations. In the same work, the investigation of the $^6$Li($^7$Li, $^6$Be)$^7$He reaction at incident energy of 450 MeV has shown for the mirror $^6$Be nucleus the analogous resonance at $E^*= 18.0\pm 1.2$ MeV with a $\Gamma$ width of $9.2 \pm 1.0$ MeV by the $^6$Be$^*$ decay into the $\tau$+$\tau$ channel. On the other hand, in the deuteron inclusive energy spectra obtained  by investigation of  the $^7$Li(n,d)$^6$He  reaction \cite{Brady77}  the $^6$He states at 0.0 and 1.8 MeV of excitation energy were  observed and  evidence of excited states at 13.6, 15.4 and 17.7 MeV was found. Therefore, the experimental and theoretical studies on the $^6$He$^*$ level distribution are very interesting because this nucleus: i) at low excitation energies appears  made of an $\alpha$ particle core with a halo of two neutrons, ii) at high excitation energies appears  constituted of two t+t clusters. Besides, the comparison between the distribution of levels for the two $^6$He and $^6$Be mirror nuclei is interesting. 

With this state of affairs, we decided to investigate other  three-body reactions such as $^3$H($\alpha$, tt)p and $^3$H($\alpha$, pt)t at $E_\alpha$  incident energy suitable to populate the  $^6$He$^*$ levels up to around 18-19 MeV of excitation energy. By these reactions we obtained ($E_{\rm t}$,~$E_{\rm t}$) and  ($E_{\rm p}$,~$E_{\rm t}$)  bidimensional spectra useful to give information on the peak energy of $^6$He$^*$ level of our interest,  also taking  into account the possible contributions in the spectra of the $^4$He$^*$ states which decay into the p+t channel for  which  the threshold energy is 19.82 MeV. In fact, as we will explain in Sect. III, along the kinematic loci of each ($E_{\rm t}$,~$E_{\rm t}$) or  ($E_{\rm p}$,~ $E_{\rm t}$) bidimensional spectrum of the above-mentioned three-body reactions,  contributions of both $^6$He$^*$  and $^4$He$^*$ state formation are present, and we have to consider that in the analysis of the t-t or p-t coincidence events.

\section{experimental set-up and coincidence event procedure}

In order to study the $^3$H($\alpha$,tt)p and $^3$H($\alpha$,pt)t reaction mechanisms by the analysis of  ($E_{\rm t}$,~$E_{\rm t}$) and  ($E_{\rm p}$,~ $E_{\rm t}$)  bidimensional spectra, we used the apparatus scheme  described in our previous work \cite{pov11} where the target made of  titanium backing (2.6 mg/cm$^2$ thick) saturated with tritium (equivalent to the thickness of about 0.15 mg/cm$^2$) and the $\alpha$-particle beam of $67.2 \pm 0.4$ MeV, produced by the isochronous cyclotron accelerator U-240 of the Institute for Nuclear Research at Kiev, were used. 

\begin{figure}[htb]
\vspace{-0.5cm}
\includegraphics[scale=0.48]{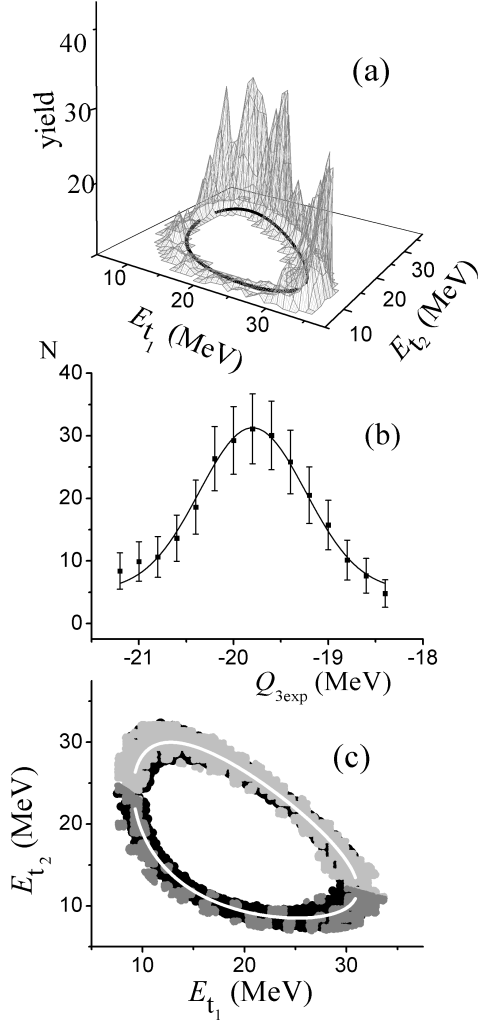} 
\vspace{-0.5cm}
\caption{(a) Experimental bidimensional  spectrum of the t-t coincidence events for the $^3$H ($\alpha$,tt)p reaction at $E_\alpha$=67.2 MeV, $\theta_1=+20^\circ$ (right side), and $\theta_2=-21^\circ$  (left side). Black solid lines represent kinematic curves for the corresponding experimental conditions. (b) Experimental Q-value distribution for the three-body reaction obtained by the bidimensional spectrum analysis;  solid line is the result of the fit. (c)  ($E_{\rm t_1}$, $E_{\rm t_2}$) bidimensional  spectrum separated in upper (light grey) and lower (grey) branches, while a black background represents Monte Carlo kinematic calculations, and  white solid lines represent kinematic calculations in the frame of a punctual geometry.}
\label{fig2}
\end{figure}

\begin{figure}[h]
\includegraphics[scale=0.5]{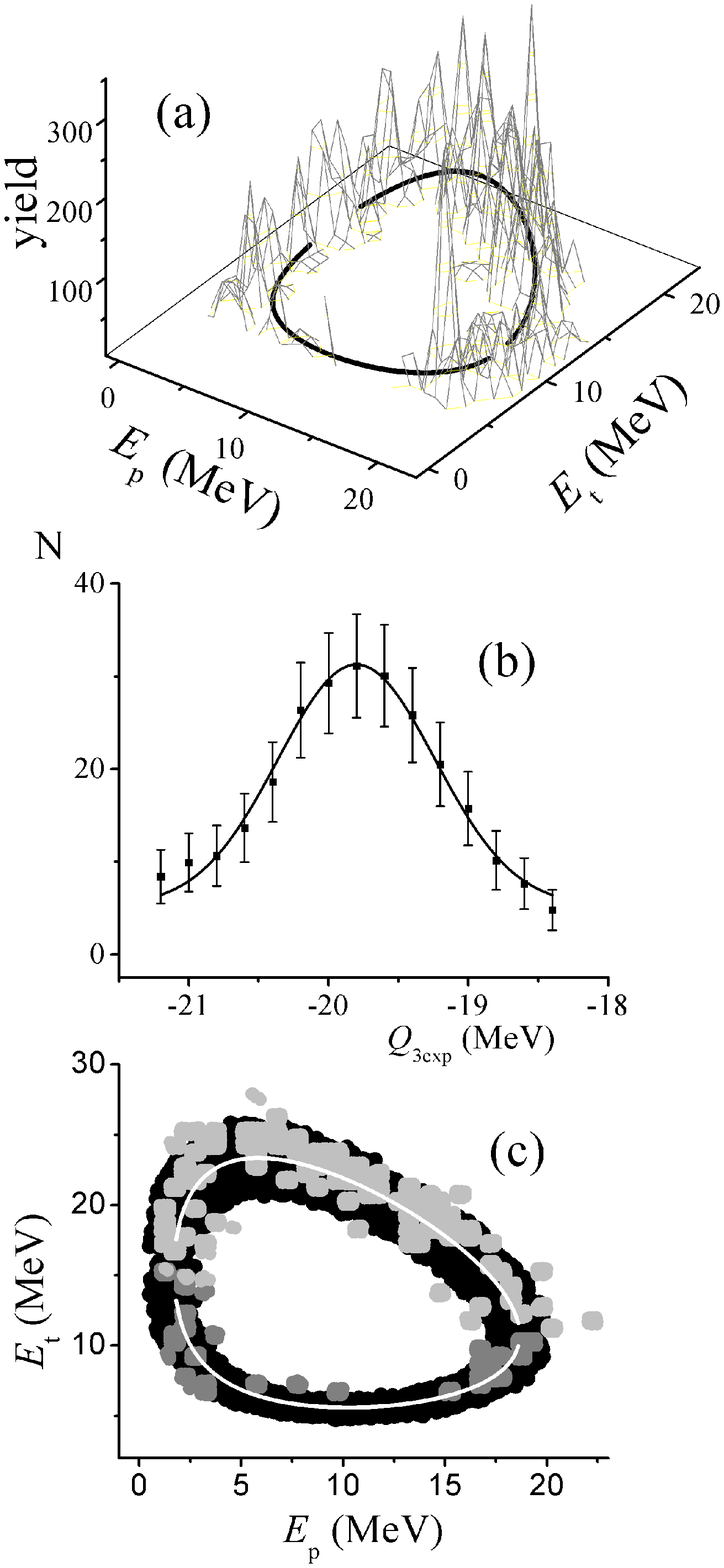} 
%\vspace{-11cm}
\caption{As Fig.~\ref{fig2}, but with detectors placed at  $\theta_p=-27.5^\circ$ (left side)  and $\theta_t=+15^\circ$(right side) for the $^3$H($\alpha$,pt)t reaction and the ($E_{\rm p}$,~$E_{\rm t}$) bidimensional spectrum.}
\label{fig3}
\end{figure}

To detect the products of the $\alpha$+t reaction and to avoid the coincidence events related to the particles present in the above-mentioned reaction that are not of our interest, we used two $\Delta E-E$ telescopes placed to the left and, to the right  of the beam direction assumed as polar axis. We used a pair of $\Delta$E-E telescopes to detect t-t  and p-t coincidences   from the $^3$H($\alpha$,ttp) reaction. The telescope placed on the right side consisted of $\Delta$E (90 $\mu$m thick totally depleted silicon surface barrier detector (SSD)) and E [Si(Li) with 3\,mm$^{t}$] detectors, while the telescope
 placed on the left side consisted of $\Delta$E [400 $\mu$m  SSD]  and E [NaI(Tl) with 20 mm$^{\phi}\times$ 20 mm$^{t}$] detectors. 
The calibration of the scintillator was made using the same procedure described in our previous paper \cite{pov11}, while a standard technique was used for the SSD. We  recorded the signals coming from the two telescopes within a window time of about 100 ns by using a standard electronic set-up. The  ($E_{\rm t}$~,~$E_{\rm t}$) and ($E_{\rm p}$~,~$E_{\rm t}$) bidimensional spectra were obtained by the t-t and p-t coincidence events  and some results are presented in Figs.~\ref{fig2}~(a) and \ref{fig3}~(a), respectively.

The corresponding experimental $Q_{\rm 3exp}$-value distributions (see Figs.~\ref{fig2}~(b) and \ref{fig3}~(b)) deduced  from the considered spectra (Figs.~\ref{fig2}~(a) and  \ref{fig3}~(a)), by using the energy and momentum conservation laws\cite{Rae84}, provide the opportunity of estimating the correctness of the related measurements, and of determining the total experimental error.  We obtain for the t-t coincidence events the experimental Q-value peak of - 19.71  MeV for the $Q_{\rm (three-body)}$  distribution (while the theoretical Q-value is -19.81  MeV) and the FWHM value of about 1.54 MeV (see Fig.~\ref{fig2}~(b)) with a standard deviation $\sigma$ of 0.65 MeV for the fit by a Gaussian function. In the case of p-t coincidences we obtain the value of - 19.80  MeV for the  experimental Q-value peak (see Fig.~\ref{fig3}~(b)) and the FWHM value of 1.32 MeV with a standard deviation of 0.56 MeV. 
Both of the $Q_{\rm (three-body)}$ experimental determinations are consistent and in agreement with the theoretical three-body $Q$-value. These results indicate the correctness of all experimental treatment and analysis procedure in the $^3$H($\alpha$, tt)p and $^3$H($\alpha$, pt)t three-body experiments
taking into account the detector resolution, beam resolution, energy straggling in the target, effect of differential target thickness, kinematic changing from beam spot size, and beam divergence.
For a further analysis of the experimental data coming from the $^3$H($\alpha$,tt)p and $^3$H($\alpha$,pt)t reactions we projected the upper and lower loci of the kinematical curves on the $E_{\rm t}$  (or $E_{\rm p}$) energy axis of tritons (or protons). As Figs.~\ref{fig2}~(c)  and \ref{fig3}(c) show, the ($E_{\rm t}$,~$E_{\rm t}$) and ($E_{\rm p}$,~$E_{\rm t}$) bidimensional  spectra are separated into upper and lower branches. By using the Monte Carlo calculation, described in \cite{pov07}  and previously used in the study of excited $^6$Li levels by the  $^3$H($\alpha$,$\tau$t)n three-body reaction \cite{pov11}, we reproduced the bidimensional coincidence event distribution obtained in the experiment by simulation. By projecting the event distribution as obtained in Fig. \ref{fig2} on the $E_{\rm t_1}$-axis for the ($E_{\rm t_1}$, $E_{\rm t_2}$) bidimensional  spectrum, or on the $E_{\rm p}$-axis for the ($E_{\rm p}$, $E_{\rm t}$) spectrum obtained in Fig. \ref{fig3}, we  analyze the various resonance contributions.

The yield of a three-body reaction, where two-body resonances at intermediate step of the process are formed, can be calculated by a sum of the Breit-Wigner contributions

  \begin{eqnarray}
   N \propto \rho(\Omega_{t_1}, \Omega_{t_2}, E_{t_1})& \times &(\sum^n_{j=1} C_j \frac{(1/2\Gamma_j)^2}{{(E_{j}-E_{pt})}^2 + {(1/2\Gamma_j)}^2} \nonumber \\
   & +  &\sum^m_{l=1} A_l \frac{(1/2\Gamma_l)^2}{{(E_{l}-E_{\rm tt})}^2 + {(1/2\Gamma_l)}^2}),
   \label{eq8}
\end{eqnarray}

where  $\rho$ is the  phase space factor of the three-body reaction, $C_{ j}$  is the corresponding contribution of each unbound $^4$He$^*$ state decaying  into the p+t particles, and  $A_l$ is the corresponding contribution of each  $^6$He$^*$ level  decaying  into the t$_1$+t$_2$ clusters. 
The values ​​of relative energies and phase space factor of the sequential three-body reaction used in expression (\ref{eq8}) are calculated  by  the Monte Carlo simulation taking into account the geometry and energy parameters of  the experiment. 

\section{Data analysis}

\begin{figure}[h]
\includegraphics[scale=0.5]{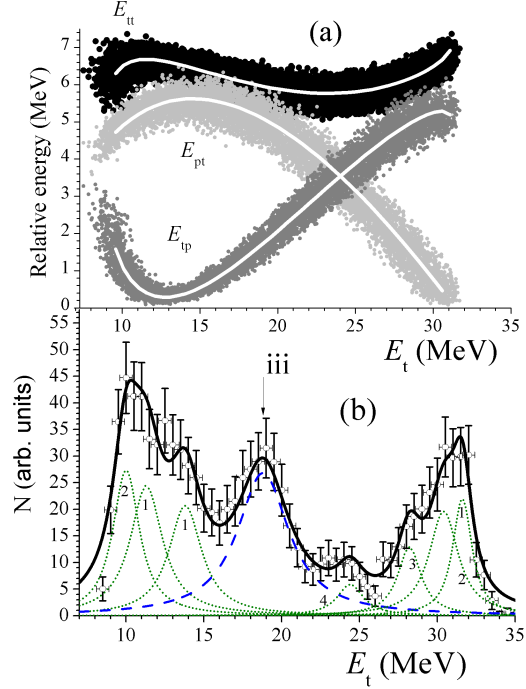}
\vspace{-0.5cm} 
\caption{(Color online)  (a) The white solid lines represent the $E_{\rm tt}$, $E_{\rm pt}$,  and $E_{\rm tp}$ relative kinetic energies  of the outgoing particle pairs versus the $E_{\rm t}$ energy for the $^3$H($\alpha$, tt)p three-body reaction,  at incident energy  $E_\alpha$ = 67.2 MeV, and with the detectors placed at $\theta_{\rm t_1}=\theta_{\rm 1} = +20^\circ$ and  $\theta_{\rm t_2}=\theta_{\rm 2} = -21^\circ$. The Monte Carlo calculations are presented as colorful arrays of dots. (b) The spectrum obtained as projection of the upper branch of the t-t coincidence yields onto the $E_{\rm t}$-axis of  detector placed at $\theta_{\rm 1}$. The central peak (dashed line) labelled as iii is due to the population of the high-lying $^6$He$^*$  state at about 18 MeV of excitation energy decaying into the t+t channel. The other resonant contributions are due to the $^4$He$^*$ state formation decaying into the p+t  particles. The ``left'' resonance structure is due to the first two  $^4$He$^*$ excited states at 20.2 and 21.0 MeV (dotted lines, labelled  as 1 and 2, respectively) and  the ``right'' resonance is due to the first four $^4$He$^*$ excited states (dotted lines, labelled  as 1, 2, 3 and 4) where the third and fourth excited states are at 21.8 and  23.3 MeV, respectively . The solid line is the sum of all contributions.}
\label{fig4}
\end{figure}

\begin{figure}[h]
\vspace{-0.5cm}
\includegraphics[scale=0.5]{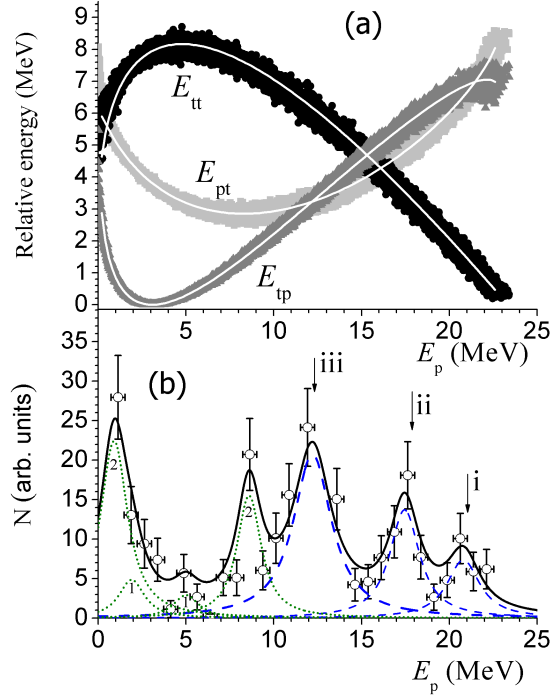}
\vspace{-0.5cm} 
\caption{(Color online)(a) The  $E_{\rm tt}$, $E_{\rm pt}$,  and $E_{\rm tp}$ relative energies of the outgoing particle pairs  for the $^3$H($\alpha$,pt)t three-body reaction at $\theta_{\rm p}=-21^\circ$ and $\theta_{\rm t}=+15^\circ$ versus the $E_{\rm p}$ energy, calculated in the frame of a punctual geometry are indicated by  white solid lines. The  Monte-Carlo calculation is presented as colorful arrays of dots. (b) The spectrum obtained as projection of the upper branch  of the p-t coincidence yields onto the $E_{\rm p}$-axis. The  peaks at $E_{\rm p}>$9.5 MeV labelled as i and ii (thin dashed lines), and iii (thick dashed line)  are the contributions  due to the population of the  14, 16 and 18  MeV $^6$He$^*$ states, respectively. The four contributions at $E_{\rm p}<$ 9.5 MeV  labelled as 1 and 2 (dotted lines) are caused by the formation of the first two $^4$He excited states at 20.2 and 21.0 MeV, respectively. The solid line is the sum of  all contributions. }
\label{fig5}
\end{figure}

\begin{figure}[h]
\includegraphics[scale=0.5]{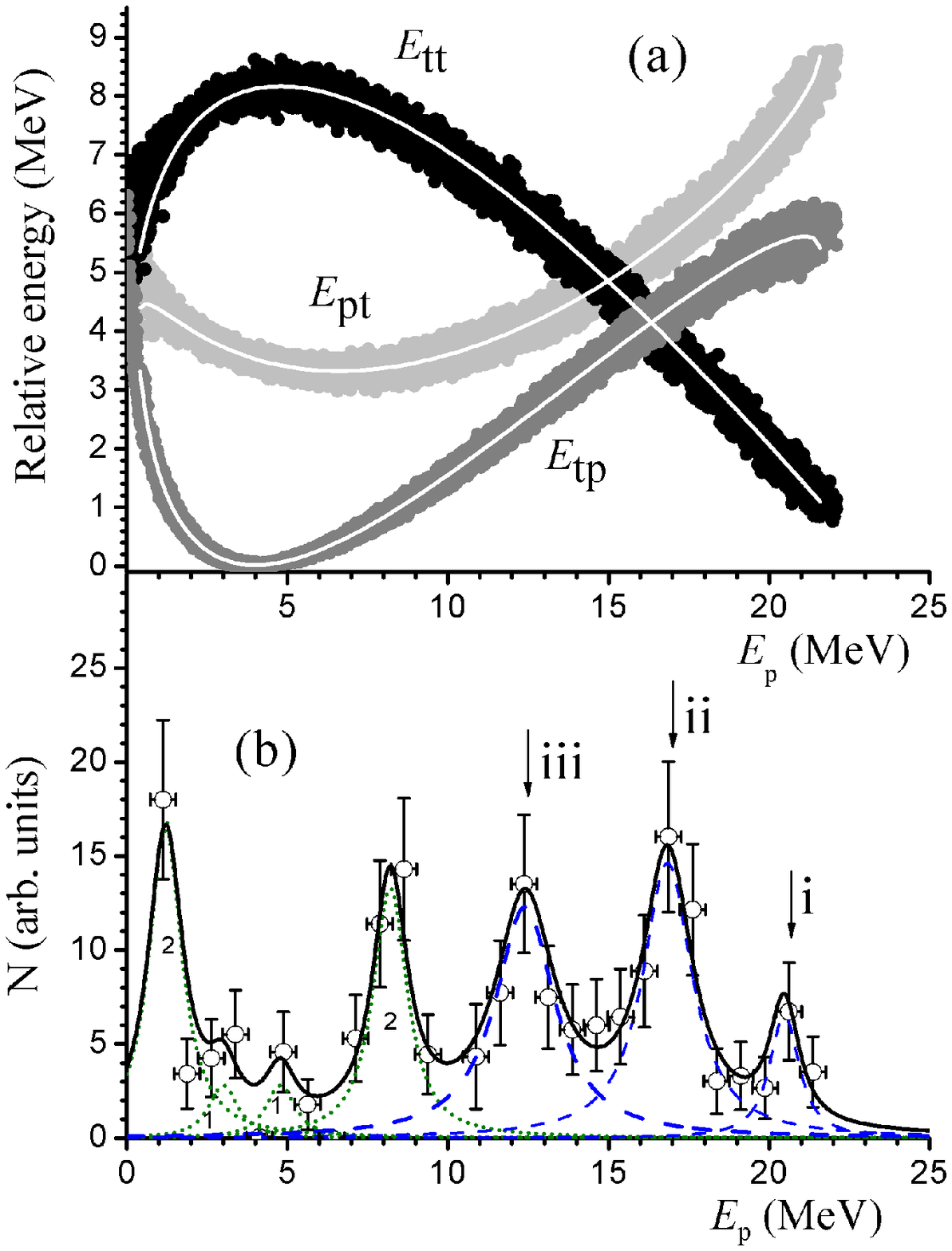} 
\vspace{-0.5cm}
\caption{(Color online) As Fig. \ref{fig5}, but for $\theta_t$=+20$^\circ$ and with the contributions of  $^4$He$^*$ state formation (panel(b)) at $E_{\rm p}<$ 10.~MeV.}
\label{fig6}
\end{figure}

The  bidimensional spectra of the  t-t and p-t coincidence events obtained by the $^3$H($\alpha$, tt)p and $^3$H($\alpha$, pt)t three-body reactions  contain experimental information about the unbound excited states of $^6$He$^*$ and $^4$He$^*$ corresponding to the t+t and  p+t systems, respectively.
Starting from the $\alpha$+$^3$H interaction in the entrance channel, the ways of forming the t+t+p products in the exit channel are the following:

  \begin{eqnarray}
\alpha  + ^3{\rm H} & \rightarrow & t + ^4{\rm He}^* \rightarrow  t + p + t  \label{eq1} \\
              & \rightarrow  & p + ^6{\rm He}^*     \rightarrow   p + t + t \label{eq2} \\
             & \rightarrow  & p +{\rm quasifree}\,\, t+t \,\,{\rm scattering} \label{eq3} \\
             & \rightarrow &  t +t+ p \label{eq4}              
\end{eqnarray}

where the (\ref{eq1}) and (\ref{eq2}) processes are the mechanisms in which   unbound states of $^4$He$^*$ and $^6$He$^*$ are formed, respectively, and then they decay into the  corresponding pairs of particles. Process (\ref{eq3}) is the quasifree t+t scattering in which the $^3$H-particle comes from  the virtual decay of  $\alpha\to$p+t.  Process (\ref{eq4}) is the statistical three-body break-up. The yield of each process depends on the kinematic conditions of reacting nuclei and geometric configuration of detectors. Therefore, we have to select the  detector angles in order to find the optimal conditions where the  $^6$He$^*$  states with the t+t cluster structure are significantly excited and the $^4$He$^*$ resonance contributions are not strongly overlapped with those of  $^6$He$^*$. In fact, we have to note that  in the case of detecting t-t coincidence particles we are not sure that all registered coincidence events located along the  ($E_{\rm 1}$, $E_{\rm 2}$) bidimensional spectrum at $\theta_1$ and $\theta_2$ detector angles correspond to events of the $^6$He$^*$ excited state formation caused only by the process (\ref{eq2}). This is because  in this bidimensional spectrum t-t coincidence events   caused by the process (\ref{eq1}) where $^4$He$^*$  states are formed are also present. Therefore, when we analyze the energy spectrum (see for example Fig.~\ref{fig4}~(b))  by projecting  the coincidence events of the upper branch of the bidimensional loci onto the $E_{\rm t}$-axis (for example, the $E_{\rm t}$ energy value registered by the detector placed at $\theta_{\rm 1}=$+20$^\circ$), we have peaks contributed by coincidence events belonging to the 
$^6$He$^*$  states (formed by process (\ref{eq2})) and also to the $^4$He$^*$  states (formed by process (\ref{eq1})). In fact, if  p is the spectator particle (the residual non-resonant particle at the first step of reaction) detected in our case at $\theta_{\rm p}=-21^\circ$ and t  (detected at $\theta_{\rm t}=+15^\circ$ in Fig. \ref{fig5} or $\theta_{\rm t}=+20^\circ$ in Fig. \ref{fig6}) is one of the two t+t cluster constituting the $^6$He$^*$ resonances decaying into two tritons at the second step of reaction populating the channel (\ref{eq2}), we can observe some  resonance features of the p-t coincidence events along the kinematic loci of the ($E_{\rm p}$,~$E_{\rm t}$) bidimensional spectrum in correspondence of some particular $E_{\rm p}$ values. 
In such a case, since we analyze a kinematically complete three-body reaction by the $E_{\rm p}$, $E_{\rm t}$,  $\theta_{\rm p}$, $\phi_{\rm p}$, $\theta_{\rm t}$,  $\phi_{\rm t}$ measurements, we are able to determine  the  $E_{\rm pt}$, $E_{\rm tp}$, and $E_{\rm tt}$ relative kinetic energies for each coincidence event. Therefore, we know what t+t excited states of $^6$He$^*$ are populated in the spectrum. On the  other hand, if t is the spectator particle detected at $\theta_{\rm t}$(+15$^\circ$ or $+20^\circ$ in Fig. \ref{fig5} or \ref{fig6}, respectively) leaving the $^4$He$^*$ states and the p particle detected at $\theta_{\rm p}=-21^\circ$ is one of the two p+t particles produced by the $^4$He$^*$ decay populating the channel (\ref{eq1}), we know which $E_{\rm tp}$ relative kinetic energy values of the $^4$He resonant states can enhance the coincidence event distribution along the kinematic loci of the ($E_{\rm p}$, $E_{\rm t}$) bidimensional spectrum. An appropriate choice of the detector angle configuration avoids the possible strong overlap of $^6$He$^*$ and $^4$He$^*$ contributions in the p-t coincidence events. Therefore, only the inspection of the coincidence event contribution projected onto the  $E_{\rm p}$-axis of Figs. \ref{fig5} and \ref{fig6} in relation to the $E_{\rm pt}$, $E_{\rm tp}$, and $E_{\rm tt}$ relative kinetic energies, by using relation (\ref{eq8}) in calculation of our analysis can allow us to individualize  $^6$He$^*$ and $^4$He$^*$ resonant contributions formed by the (\ref{eq1}) and (\ref{eq2}) processes which are both present in the bidimensional ($E_{\rm p}$, $E_{\rm t}$) spectra of the p+t+t three-body reaction. We then determine the $E^*$ and $\Gamma$ width values for all $^6$He$^*$ and $^4$He$^*$ resonant contributions considered in Figs. \ref{fig4}~(b), \ref{fig5}~(b), and \ref{fig6}~(b) as results of the fit procedure.

 Therefore, without an accurate analysis of the various resonance peaks formed by the decay of both $^6$He$^*$ and $^4$He$^*$  states that we consider in this work (taking into account the $E_{\rm tt}$, $E_{\rm pt}$, and  $E_{\rm tp}$ relative energies  and various resonant contributions in formula (\ref{eq8})), it is impossible to separate  the decay contributions  of the various $^6$He$^*$ states from the ones caused by the decay of the $^4$He$^*$ states, or to determine  the  $E^*$ excitation energy and $\Gamma$ width values of the investigated $^6$He$^*$ states in a reliable way. Of course, the choice of the angular configuration of detectors and projection of events onto the $E_{t}$- or  $E_{p}$-axis in order to obtain the energy spectrum can favour the yields of some $^6$He$^*$ peaks in comparison with the ones of  $^4$He$^*$ peaks, but the competition between the  (\ref{eq1}) and (\ref{eq2}) processes and presence of both their contributions are not eliminable by any hardware or software treatment. Therefore, we need to take into account   all  possible $^6$He$^*$ and $^4$He$^*$  state contributions  in relation (\ref{eq8}) in the analysis.

We think it is important to investigate the existence of the 18 MeV $^6$He$^*$  energy region by analyzing the bidimensional spectra obtained by the $^4$He+$^3$H reaction  and to compare  our results with the ones obtained by the $^7$Li+$^6$Li \cite{Yamagata05}  and  n+$^7$Li \cite{Brady77} experiments. In order to observe the effects of the  18 MeV $^6$He$^*$ state formation by its decay into the  t+t channel we selected the  angles of telescopes as $\theta_{\rm 1}$=+20$^\circ$  and  $\theta_{\rm 2}=-$21$^\circ$  in order to make the $E_{\rm tt}$ relative energy function very flat around the excitation energy of our interest along the bidimensional kinematic curve. This is an optimal condition to determine the $E^*$ energy of the  $^6$He$^*$ state with the best energy resolution because the flat behaviour of the $E_{\rm tt}$ relative energy  avoids distortion effects due to the projection of coincidence events on the $E_{\rm t_1}$-axis (or analogously on the $E_{\rm t_2}$-axis)
of the two-dimensional event distribution. However  such mentioned detector geometry  does not allow a  better determination of the $\Gamma$ width value. In fact, in order to obtain the more realistic determination of $\Gamma$ width it is convenient to project the ($E_{\rm p}$,~$E_{\rm t}$)  coincidence events on the $E_{\rm p}$-axis, and eventually to select appropriate $\theta_{\rm p}$ and $\theta_{\rm t}$ detector angles so that it is possible to range a larger interval of $E_{\rm tt}$ relative energies around the 18 MeV $^6$He$^*$ peak energy (see Figs.~\ref{fig5} and \ref{fig6}).

   In a bidimensional spectrum, the finite angular and energy resolutions of  detectors contribute to the spreading of events in the ($E_{\rm 1}$, $E_{\rm 2}$) plane, where $E_{\rm 1}$ and $E_{\rm 2}$ are the energies of the two detected coincidence particles. Therefore, it should be necessary to separate the geometric effects from the energy ones before  the treating of data. Since this  is a  very difficult task, we  decided to analyze the  bidimensional spectra by the Monte Carlo method as we did in a  recent previous work \cite{pov11}.  We have generated a sufficient set of random events suitable to obtain the t-t or p-t coincidences. In the Monte Carlo simulation we take into account the value of the beam energy and its dispersion, the thickness of the target, the energy loss in the target, the size of the spot beam on the target, the target-detector distances, and the energy resolution of detectors. To analyze the experimental data obtained by the $^3$H($\alpha$, ttp) reaction, we should project the upper and the lower loci of the kinematic curve onto the $E_{\rm t_{1}}$  energy axis (see Fig.~\ref{fig2}) or onto the $E_{\rm p}$   energy axis (see Fig.~\ref{fig3}).    
   The procedure is performed by recalculating the  ($E_{\rm t_{1}}$, $E_{\rm t_{2}}$) or ($E_{\rm p}$,~$E_{\rm t}$) bidimensional spectra of  coincidence events by using the Monte Carlo method and projecting their yields onto the $E_{\rm t}$ or $E_{\rm p}$ axis. The selected ($E_{\rm t_{1}}$, $E_{\rm t_{2}}$) bidimensional spectrum, obtained for the $E_\alpha$ beam energy of 67.2 MeV and detectors placed at $\theta_{\rm t_{1}}=\theta_{\rm 1}=+20^\circ$ and  $\theta_{\rm t_{2}}=\theta_{\rm 2}=-21^\circ$ are divided into upper and lower branches (see Fig.~\ref{fig2}) by using the above-mentioned method, and the upper branch of bidimensional loci is projected onto  the $E_{\rm t_{1}}$ energy axis (see Fig.~\ref{fig4}~(b)). Moreover, Fig.~\ref{fig4}~(a) shows the relative kinetic energies of the t-t,  p-t, and t-p pairs of particles versus $E_{\rm t_{1}}$, where $E_{\rm t_{1}}$ is the energy value of the triton that is registered by the detector placed at $\theta_{\rm 1}=+20^\circ$ while the other triton of $E_{\rm t_{2}}$ energy is registered by the detector placed at $\theta_{\rm 2}=-21^\circ$.
The  analysis of the full resonant structures that appear in the spectrum of Fig.~\ref{fig4}~(b) joined with the corresponding $E_{\rm tt}$, $E_{\rm pt}$, and $E_{\rm tp}$ relative kinetic energies described by the lines reported in Fig.~\ref{fig4}~(a) allows us to know if one resonant peak of the event distribution is contributed by some $^6$He excited states that decay into the t+t particles, or by some  $^4$He excited states that decay into p+t particles, or eventually if the peak can be formed by some overlapped contributions caused by the decay channels of  $^6$He$^*$ and $^4$He$^*$ states.  In the fit procedure of relation (\ref{eq8}) we use the $E^*$ and $\Gamma$ width parameters given in Ref. \cite{till92} as starting values for the $^4$He$^*$ resonances giving the calculation procedure  the possibility of adjusting such parameters, while the values were fully free for the parameters of the $^6$He$^*$ resonances.

    As Fig.~\ref{fig4}~(a) shows, it is evident that the trend of the $E_{\rm tt}$ function remains almost constant with a small fluctuation around  the 6.0 MeV value.      
  Taking into account  the threshold energy of  12.31 MeV for the  $^6$He$^*$ level that decays into the t-t channel, the peak energy of the event distribution included in the 16-23 MeV $E_{\rm t}$ energy range corresponds to the excitation energy of 18.3 MeV for the $^6$He$^*$ nucleus. Therefore, this $^6$He excited state was populated in our $^3$H($\alpha$,tt)p experiment. Moreover, the figure shows that the set of  full $E_{\rm tt}$ values around the considered  $^6$He$^*$ level  ranges within the interval of about 0.7 MeV. This means that, in respect to the behaviour of the $E_{\rm tt}$ shape, the energy peak is well determined with a small error of $\pm 0.2$ MeV while the FWHM determination is affected by the small 0.7 MeV  range  of the $E_{\rm tt}$ relative energy values around the peak at 18.3 MeV of $^6$He$^*$ state. Instead, by the other analyzed energy spectra versus $E_{\rm p}$ corresponding to different detector angles we can determine the $\Gamma$ width value of the above-mentioned $^6$He$^*$  state in a reliable way (see Figs.~\ref{fig5}~(b) and \ref{fig6}(b)) because the relative energy values of the complete $E_{\rm tt}$ function range in an interval of about 8 MeV. 
  
Figure \ref{fig4}~(b) shows the event distribution due to the projection of the upper branch of the ($E_{\rm t_1 }$,$E_{\rm t_2 }$) bidimensional spectrum onto the $E_{\rm t_1}$ energy measured by the detector placed at $\theta_{\rm 1}=$20$^\circ$. The error bars take into account only the statistical error, while the  finite energy resolution of the used electronic system is about 0.4 MeV. As one can see, three resolved contributions appear in this figure.  On the left side, in the  8~MeV$<E_{\rm t}<$16~MeV energy range, the main resonance contributions  are due to the population  of the first two $^4$He  excited states and their decay into the p+t channel (process (\ref{eq1})), when the first emitted t-particle (the spectator in the process (\ref{eq1})) goes to the detector placed at $\theta_1$  while the t-particle coming from the decay of $^4$He excited states    into the p+t channel goes to the detector placed at $\theta_2$; the particular trend of the $E_{\rm t }$  line with inversion of the relative kinetic energy value of the t+p system in Fig. \ref{fig4}~(a) leads to the repetition of some $^4$He$^*$ resonant state contributions at increasing  $E_{\rm t }$ energy values along the $E_{\rm t }$-axis. 
In the central part of the figure,  in the  16-23 MeV $E_{\rm t}$ energy range,  the main contribution is due to the 18 MeV $^6$He$^*$ state formation and to its decay  into two tritons detected at $\theta_1$ and $\theta_2$ angles (process (\ref{eq2}));  on the  right side, in the 23~MeV$<E_{\rm t}<$ 35~MeV energy range, a wide complex resonant structure is due to contributions of the  first four $^4$He   excited states which decay  into the p+t channel with the first t-particle (the spectator in the process(\ref{eq1})) that goes to the detector  placed at $\theta_2$  while the t-particle coming from the decay of $^4$He excited states into the p+t channel goes to the detector  placed at $\theta_1$.

The final calculation result  obtained by using expression (\ref{eq8}) and the least squares method with variables describing the energy peak and width of the various contributions of  the $^4$He$^*$  and $^6$He$^*$  states is reported in Fig.~\ref{fig4}~(b) by a solid line. The dotted lines represent the single contributions of the various   $^4$He$^*$ levels while the dashed line represents the resonant contribution of the 18.3 MeV  $^6$He$^*$ state. 
The obtained values of  $18.3 \pm 0.2$~MeV for the $E^*$ energy peak and   $0.4 \pm 0.2$~MeV for the $\Gamma$ width regarding the mentioned $^6$He$^*$ state, labelled as iii in Fig.~\ref{fig4}~(b), are also reported in Table~\ref{tb1}.

With the aim of checking these obtained results in the 18.3 MeV $^6$He$^*$ state and in order to extend our investigation on the other near high-lying $^6$He$^*$ states, we  studied  ($E_{\rm p}$,~$E_{\rm t}$) bidimensional spectra by projecting the p-t coincidence events onto the $E_{\rm p}$-axis with different geometric configuration of detectors and by analyzing the obtained $E_{\rm p}$ energy spectra. At first we selected the coincidence events when the proton goes to the detector placed at  $\theta_{\rm p}=\theta_1=-21^\circ$ while the triton is detected at $\theta_{\rm t}=\theta_2=+15^\circ$. As Fig.~\ref{fig5}~(a) shows,  the shapes of the $E_{\rm tt}$, $E_{\rm pt}$, and $E_{\rm tp}$ relative energies  of the t-t, p-t and t-p systems are very different in comparison with ones presented in Fig.~\ref{fig4}~(a). In fact, in the case of Fig.~\ref{fig5}~(a) it is possible to explore the $E_{\rm tt}$ relative energy range  of about 8 MeV for the decay of  $^6$He$^*$ states into the t-t channel. 

Figure \ref{fig5}~(b) shows the energy spectrum of the event distribution obtained by projection of the upper branch of coincidence events versus the $E_{\rm p}$ energy measured by the detector placed at $\theta_p=-21^\circ$. In this $E_{\rm p}$  energy spectrum the resonance structures at $E_{\rm p}<$~9.5 MeV are contributed by the first two $^4$He$^*$ states, labelled in figure as 1 and 2, while the  ones at $E_{\rm p}>$9.5 MeV are contributed by the three $^6$He$^*$ levels at excitation energies included in the 13.5-19.0 MeV range. Analogously to what was observed in the description of the left part of the spectrum in Fig. \ref{fig4}, in the case of the $E_{\rm p}<9.5$~MeV region the contributions of the first two excited $^4$He state formation are present twice  at increasing  $E_{\rm p}$, due to the inversion and repetition of the $E_{tp}$ relative kinetic energy values of the t+p system.  Therefore, by using expression (\ref{eq8}), we obtain by the least squares calculation method the values of $E^*$ and $\Gamma$ parameters for the $^6$He$^*$ states, labelled in figure and Table \ref{tb1} as i, ii, and iii, corresponding to the $^6$He excitation energies of $14.0 \pm 0.4$, $15.8 \pm 0.4$,  and $18.5 \pm 0.4$ MeV with $\Gamma$ width values of $0.7 \pm 0.3$, $0.8 \pm 0.3$, and $1.1 \pm 0.4$, respectively. The present analysis of the ($E_{\rm p}$,~$E_{\rm t}$) bidimensional spectrum  confirms the population   of the  18 MeV $^6$He$^*$  level found  by the fit of data of Fig.~\ref{fig4}~(b) where  the peak energy of $18.3 \pm 0.2$~MeV was determined with  a better energy resolution, while in the analysis of the energy spectrum of Fig.~\ref{fig5}~(b) versus the $E_{\rm p}$-axis a realistic $\Gamma$ determination of $1.1\pm 0.4$~ MeV for the mentioned $^6$He  level  was obtained from a spectrum where the set of values of the $E_{\rm tt}$ relative energies  ranges in the  8 MeV interval which includes the  full values describing the complete energy spectrum of the found $18.5 \pm 0.4$~MeV  $^6$He excited state.  

\begin{widetext}
\begin{table*}
\caption{$E^*$ excitation energy and $\Gamma$ width values of the $^6$He$^*$  levels for different $\theta_1$, $ \theta_2$ geometric detector configurations, as results of  the Breit-Wigner approximation by using formula(\ref{eq8}).}\label{tb1}
\begin{tabular}{cccccc}
\hline
\hline
 $\theta_1$, $ \theta_2$ &\,\, Reaction \,\,& peak label\,\, & $E^*$ (MeV) \,\,& $\Gamma$ (MeV) & \,\, see Fig. \\  
\hline 
 +20$^\circ$, -21$^\circ$ & \,\, $^3$H($\alpha$,tt)p\,\, & iii  \,\,& 18.3 $\pm$  0.2 &\,\,\, 0.4 $\pm$  0.2\footnote{this  $\Gamma$ value obtained in the analysis of the energy spectrum of Fig.~ \ref{fig4}~(b) does not correspond to the true $\Gamma$ width of the 18.3 MeV $^6$He$^*$ state  because its determination by the ($E_{\rm t}$,~$E_{\rm t}$) spectrum is affected by the limited accessible $E_{\rm tt}$ relative energy interval of about 0.7 MeV only (see Fig.~\ref{fig4}~(a) and text for details).} &\,\, \ref{fig4}~(b) \\
 -21$^\circ$, +15$^\circ$ & \,\,$^3$H($\alpha$,pt)t\,\, & i  \,\,& 14.0 $\pm$  0.4 & \,\, 0.7 $\pm$  0.3 &\,\, \ref{fig5} (b) \\
                           & \,\,                  & ii \,\,& 15.8 $\pm$  0.4 & \,\, 0.8 $\pm$  0.3&\,\, '' \\
                          & \,                   & iii\,\,\,& 18.5 $\pm$  0.4 & \,\, 1.1 $\pm$  0.3 &\,\, '' \\                  
 -21$^\circ$, +20$^\circ$ & \,\,$^3$H($\alpha$,pt)t\,\, & i  \,\,& 14.0 $\pm$  0.4 & \,\, 0.6 $\pm$  0.4 &\,\, \ref{fig6} (b) \\
                         & \,\,                     & ii \,\,& 16.1 $\pm$  0.4 & \,\, 0.8 $\pm$  0.4 &\,\, '' \\
                        &  \,                     &iii\,\,\,& 18.4 $\pm$  0.4 & \,\, 1.0 $\pm$  0.4 &\,\, '' \\
 \hline
 \hline
\end{tabular}
\end{table*}
\end{widetext}

Analogously to what is described in Fig.~\ref{fig5}, in Fig.~\ref{fig6} we report  the results of the analysis results of the ($E_{\rm p}$,~$E_{\rm t}$) bidimensional spectrum  obtained for detectors placed at $\theta_p=-21^{\circ}$ and $\theta_t=+20^{\circ}$.   
Figures \ref{fig6}~(a) and \ref{fig6}~(b) show  results similar to the ones analyzed in Figs.~\ref{fig5}~(a) and \ref{fig5}~(b), respectively. The four peaks present in the  $E_{\rm p}<$ 10 MeV energy range are caused by the contributions of the first two excited $^4$He states, as clearly appears by observing the inverse trend of the $E_{\rm tp}$ relative energy values around  $E_{\rm p}$=4 MeV of Fig.~\ref{fig6}~(a). The three peaks present in the $E_{\rm p}>$ 10 MeV energy range are  caused by the contributions of  $^6$He$^*$ states at excitation energies of $14.0 \pm 0.4$, $16.1 \pm 0.4$, and $18.4 \pm 0.4$, respectively.
  The results of $E^*$ and $\Gamma$ parameters related to the $^6$He$^*$ states, labelled in Fig.~\ref{fig6}~(b) as i, ii, and iii, are reported in Table~\ref{tb1}.
  
  In our analysis and fit results of spectra presented in Figs.~\ref{fig4}~(b), \ref{fig5}~(b), and \ref{fig6}~(b), for all peaks contributed by the considered $^4$He excited states  
$E^*$ and $\Gamma$ values consistent with the ones given in literature \cite{till92} were found. 
   
\begin{figure}[h]
  \vspace{3.5cm}
\includegraphics[scale=0.34]{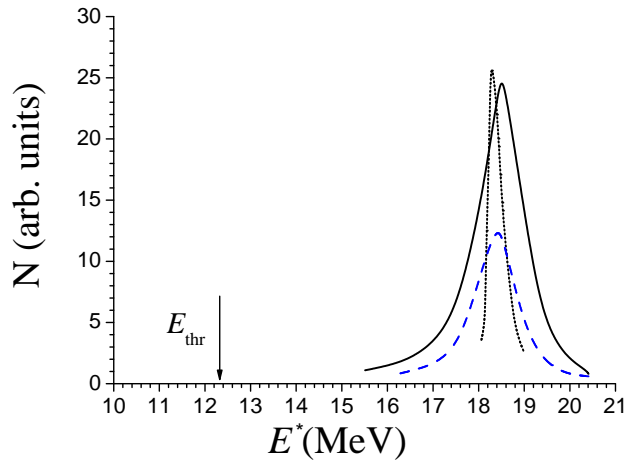} 
  \vspace{-3.0cm}
\caption{(Color online) Energy spectrum distribution of the 18.3 MeV  $^6$He level as a function of the $E^*$ excitation energy of the $^6$He nucleus. Dotted line is obtained from the $^6$He$^*$ peak contribution around the  energy interval  centered at $E_{\rm t}=$~18.8 MeV of Fig.~\ref{fig4} (b); full line is obtained from the $^6$He$^*$ peak contribution  centered at $E_{\rm p}=$12.2 MeV of Fig.~\ref{fig5} (b); dashed line is obtained from the peak at $E_{\rm p}=$12.4 MeV of Fig. \ref{fig6}~(b). The arrow indicates  the threshold energy of 12.31~MeV for $^6$He$^*$ decaying into the t+t channel. }
\label{fig7}
\end{figure}

    By considering the single contribution of the  18.3 MeV $^6$He$^*$ state formation obtained in Fig.~\ref{fig4}~(b) in the analysis of the ($E_{\rm t}$,~$E_{\rm t}$) energy spectrum, and also considering the energy distributions of the analogous $^6$He$^*$ state obtained in the analysis of the  ($E_{\rm p}$,~$E_{\rm t}$) energy spectra  given  in Figs.~\ref{fig5}~(b) and \ref{fig6}~(b),  in Fig.~\ref{fig7} we present the energy spectrum distribution of the mentioned $^6$He$^*$ state as a function of the $E^*$ excitation energy of the $^6$He nucleus. The full and dashed lines, corresponding to the  18 MeV $^6$He$^*$ state  represented in Figs.~\ref{fig5}~(b) and \ref{fig6}~(b), respectively,  show the same results for the $E^*$ energy peak and $\Gamma$ width values for  the investigated ($E_{\rm p}$,~$E_{\rm t}$) bidimensional spectra; the dotted line, corresponding to the  18.3 MeV $^6$He$^*$ state  represented in Fig.~\ref{fig4}~(b),   shows the same 
  $E^*$ peak value but a limited $\Gamma$ width value of 0.4 MeV since it is  affected by the incomplete set of the   $E_{\rm tt}$ relative energy values reached in the analyzed ($E_{\rm t}$,~$E_{\rm t}$) bidimensional spectrum. Therefore, the choice of using the t-t coincidence events and detector angles at $+20^\circ$ and $-21^\circ$ leads to a more reliable condition for the $E^*$ peak determination of $18.3\pm0.2$~MeV, while its observed $\Gamma$ width value of $0.4\pm0.2$~MeV is limited to cause the partial accessible $E_{\rm tt}$ relative energy interval of about 0.7 MeV only, instead of the 8 MeV  $E_{\rm tt}$ interval that is explored in the analyzed spectra of the  p-t coincidence events. Consequently, the $\Gamma$ values are correctly determined by the analysis of the ($E_{\rm p}$,~$E_{\rm t}$) spectra. Nevertheless, in Table~\ref{tb1} we also report the $\Gamma$ width value determined by the analysis of the  ($E_{\rm t}$,~$E_{\rm t}$) bidimensional spectrum only to understand the reasons why  the analysis of this spectrum  leads to a smaller $\Gamma$ value. In a practical way it is impossible to compare this $\Gamma$ width value of 0.4 MeV  extracted by the  ($E_{\rm t}$,~$E_{\rm t}$) energy spectrum  of Fig.~\ref{fig4}~(b) with the $\Gamma$ width values obtained from ($E_{\rm p}$,~$E_{\rm t}$) energy spectra of Figs.~\ref{fig5}~(b) and  \ref{fig6}~(b).

  In literature not many results about the 18 MeV $^6$He$^*$ state with the determination of  $E^*$ and $\Gamma$ parameters, and their respective errors are present. Brady {\it et al.} \cite{Brady77} in the  $^7$Li(n, d)$^6$He experiment at $E_n$ = 56.3 MeV observed (beside the ground state and the 1.8 MeV $^6$He$^*$ excited state) a group of at least three excited states centered near 13.6, 15.4 and 17.7 MeV  by the analysis of deuteron angular distributions. However, due to poor resolution, limited statistical accuracy, uncertainty in the energy width of these states, the authors combined these three experimental states as a single broad peak centered at 15.6 MeV.
  Moreover, Yamagata, Akimune  {\it et al.} \cite{Yamagata05} found in their  $^6$Li($^7$Li, $^7$Be t)$^3$H experiment, at $E_{\,^7{\rm Li}}$=455 MeV by the analysis of the t-particle single spectra, the resonance of $^6$He$^*$ state decaying into the t+t  clusters at $E^* = 18 \pm 1$~MeV with a $\Gamma$ width of $9.5 \pm 1$~MeV (see Fig.~3~(c) of Ref. \cite{Yamagata05}).

 We believe that in the analyzed single spectra of Ref. \cite{Brady77} it was impossible to solve the various resonance contributions of excited states present in the region of  $^6$He levels included in the range between the threshold energy of 12.31 MeV for the decay into the t+t channel and the excitation energy of about 22 MeV. 
 
 Also in the  Akimune {\it et al.} \cite{Akimune03} and  Yamagata {\it et al.}'s experiments \cite{Yamagata05}, by the analysis of the t-particle single spectra, it was  impossible  to solve  the true t-t resonant contribution of the $^6$He excited state at about $E^*=18$~MeV due to the various high-lying $^6$He levels populated by  channel (\ref{eq2}) from the contributions caused by  channel (\ref{eq1}). This is where the first t-particle leaves the various $^4$He$^*$ state formations while the second t-particle comes from the decay  of $^4$He$^*$ into the p+t system. In addition, the background contribution of the t-particles caused by the (\ref{eq3}) and  (\ref{eq4}) channels is also present in the collected spectrum. Therefore, the resulting analyzed spectrum of the authors \cite{Akimune03,Yamagata05}  appears as a convolution of the various  resonant contributions caused by the (\ref{eq1}) and  (\ref{eq2}) channels in addition to the non resonant  background due to the  (\ref{eq3}) and  (\ref{eq4}) channels. Instead, in  the analysis of the ($E_{\rm t}$,~$E_{\rm t}$) and ($E_{\rm p}$,~$E_{\rm t}$) bidimensional spectra of our $^3$H($\alpha$,ttp) kinematically complete experiment, it  was possible to separate and  observe the population of  the 18.3 MeV $^6$He$^*$ state from the other resonant contributions and also to find reliable $\Gamma$ width values for the mentioned $^6$He$^*$ state. Moreover, we also found the $\Gamma$ width values of 0.6$\pm 0.4$ - 0.7 $\pm 0.3$ MeV for the 14.0 $\pm 0.4$ MeV $^6$He$^*$ state and 0.8 $\pm 0.4$  MeV for the 16.1 $\pm 0.4$ MeV $^6$He$^*$ state. These last $\Gamma$ width determinations for the 14.0  and 16.1 MeV $^6$He$^*$ states are new findings because such found values are completely  different from the ones reported in compilations \cite{selov88,till02}.

 \section{Conclusions}
 
 In our described $^4$He+$^3$H experiment, at $E_\alpha$ beam energy of 67.2 MeV,    the $^6$He$^*$  level at $18.3 \pm 0.2$~MeV was  
 populated. This level shows the $\Gamma$ width of  $1.1 \pm 0.4$~MeV  by its decay into the binary t+t channel. These  results have been obtained by analyzing the ($E_{\rm t}$,~$E_{\rm t}$) and  ($E_{\rm p}$,~$E_{\rm t}$) bidimensional spectra at various detector geometry angles  by reactions leading to the p+t+t three-body system through the (\ref{eq1}) and (\ref{eq2}) processes where $^4$He$^*$ and $^6$He$^*$ resonant states are formed, respectively . The study shows 
resonant contributions due to the population of  high-lying  $^6$He$^*$ levels by projecting the coincidence events either onto the $E_{\rm t}$-axis or onto  the $E_{\rm p}$-axis. In both cases, the observed peaks in the energy spectra of coincidence events confirm the population of the $^6$He$^*$ state at $E^*$ of 18.3 MeV by the analysis of the  $E_{\rm tt}$ relative energies of coincidence events (see Figs.~\ref{fig4}~(b), \ref{fig5}~(b), and \ref{fig6}~(b)). In the considered energy spectra   peaks formed by the coincidence events are also present  due to the decay of the  $^4$He$^*$ state formation into the p+t channel by the analysis of the $E_{\rm pt}$ and $E_{\rm tp}$ relative energies. The reliability of the analysis of spectra and results obtained in our experiment is confirmed by the experimental $Q$-value distributions shown in Figs.~\ref{fig2}~(b) and \ref{fig3}~(b), by the clear evidence of both contributions of $^6$He$^*$ and $^4$He$^*$ states in Figs. \ref{fig4}~(b), \ref{fig5}~(b) and \ref{fig6}~(b), and by the complete consistent results shown in Fig.~\ref{fig7} and Table~\ref{tb1}.
 In the present experiment we also observed the population of  two other $^6$He$^*$  states at excitation energy $E^*$ at 14.0 and 16.1 MeV. The $E^*$ energy peak values are in agreement with the ones reported in  \cite{selov88} and \cite{till02}. On the contrary, the found $\Gamma$ width values are strongly different from the ones reported in the above mentioned compilations.  We also compared our  $E^*$ and $\Gamma$ results about the 18.3 MeV $^6$He$^*$ state with the ones obtained by  Yamagata {\it et al.} \cite{Yamagata05}, Akimune {\it et al.} \cite{Akimune03}, and  Brady  {\it et al.} \cite{Brady77} in their experiments by the $^6$Li($^7$Li, $^6$He)$^7$Be \cite{Yamagata05,Akimune03} and $^7$Li(n, d)$^6$He  \cite{Brady77} reactions, and discussed     the relevant difference in the 
  $\Gamma$ width value determinations.

The choice of the beam  energy, detector geometry, and kind of reaction   allowed us to measure the peak energy of the high-lying 18.3 MeV $^6$He$^*$ level with the  precision of 0.2 and 0.4 MeV in the energy spectra   of the $^3$H ($\alpha$, tt)p and $^3$H($\alpha$, pt)t reactions, respectively, where for this level the $\Gamma$ width determinations were 1.1 $\pm 0.3$ and 1.0 $\pm 0.4$ MeV (see Fig. \ref{fig7} and Table \ref{tb1}) by the two analyzed ($E_{\rm p}$,~$E_{\rm t}$) energy spectra. Moreover, 
we explained the reasons which led to some relevant differences among the results of the  investigated reactions, since in each observed ($E_{\rm t}$,~$E_{\rm t}$) or  ($E_{\rm p}$,~$E_{\rm t}$) energy spectrum both contributions of $^6$He$^*$ and $^4$He$^*$ excited states are present. Moreover, we also found realistic $\Gamma$ width values of about $0.7\pm~0.3$ and $0.8\pm~0.3$ MeV for the 14.0 and 16.1 MeV $^6$He$^*$ levels, respectively.

\begin{acknowledgments}
The authors wish to thank the staff of the Institute for Nuclear Research Laboratories (Kiev) for their help during the measurements. This work was supported by INR of the Academy of Sciences of Ukraine, and partially by the Istituto Nazionale di Fisica Nucleare of Italy. 
\end{acknowledgments}

% Create the reference section using BibTex:
\bibliography{povbase}

\providecommand{\noopsort}[1]{}\providecommand{\singleletter}[1]{#1}%
\begin{thebibliography}{9}%
\makeatletter
\providecommand \@ifxundefined [1]{%
 \@ifx{#1\undefined}
}%
\providecommand \@ifnum [1]{%
 \ifnum #1\expandafter \@firstoftwo
 \else \expandafter \@secondoftwo
 \fi
}%
\providecommand \@ifx [1]{%
 \ifx #1\expandafter \@firstoftwo
 \else \expandafter \@secondoftwo
 \fi
}%
\providecommand \natexlab [1]{#1}%
\providecommand \enquote  [1]{``#1''}%
\providecommand \bibnamefont  [1]{#1}%
\providecommand \bibfnamefont [1]{#1}%
\providecommand \citenamefont [1]{#1}%
\providecommand \href@noop [0]{\@secondoftwo}%
\providecommand \href [0]{\begingroup \@sanitize@url \@href}%
\providecommand \@href[1]{\@@startlink{#1}\@@href}%
\providecommand \@@href[1]{\endgroup#1\@@endlink}%
\providecommand \@sanitize@url [0]{\catcode `\\12\catcode `\$12\catcode
  `\&12\catcode `\#12\catcode `\^12\catcode `\_12\catcode `\%12\relax}%
\providecommand \@@startlink[1]{}%
\providecommand \@@endlink[0]{}%
\providecommand \url  [0]{\begingroup\@sanitize@url \@url }%
\providecommand \@url [1]{\endgroup\@href {#1}{\urlprefix }}%
\providecommand \urlprefix  [0]{URL }%
\providecommand \Eprint [0]{\href }%
\providecommand \doibase [0]{http://dx.doi.org/}%
\providecommand \selectlanguage [0]{\@gobble}%
\providecommand \bibinfo  [0]{\@secondoftwo}%
\providecommand \bibfield  [0]{\@secondoftwo}%
\providecommand \translation [1]{[#1]}%
\providecommand \BibitemOpen [0]{}%
\providecommand \bibitemStop [0]{}%
\providecommand \bibitemNoStop [0]{.\EOS\space}%
\providecommand \EOS [0]{\spacefactor3000\relax}%
\providecommand \BibitemShut  [1]{\csname bibitem#1\endcsname}%
\let\auto@bib@innerbib\@empty
%</preamble>
\bibitem [{\citenamefont {Povoroznyk}\ \emph {et~al.}(2011)\citenamefont
  {Povoroznyk}, \citenamefont {Gorpinich}, \citenamefont {Jachmejov},
  \citenamefont {Mokhnach}, \citenamefont {Ponkratenko}, \citenamefont
  {Mandaglio}, \citenamefont {Curciarello}, \citenamefont {Leo}, \citenamefont
  {Fazio},\ and\ \citenamefont {Giardina}}]{pov11}%
  \BibitemOpen
  \bibfield  {author} {\bibinfo {author} {\bibfnamefont {O.}~\bibnamefont
  {Povoroznyk}}, \bibinfo {author} {\bibfnamefont {O.~K.}\ \bibnamefont
  {Gorpinich}}, \bibinfo {author} {\bibfnamefont {O.~O.}\ \bibnamefont
  {Jachmejov}}, \bibinfo {author} {\bibfnamefont {H.~V.}\ \bibnamefont
  {Mokhnach}}, \bibinfo {author} {\bibfnamefont {O.}~\bibnamefont
  {Ponkratenko}}, \bibinfo {author} {\bibfnamefont {G.}~\bibnamefont
  {Mandaglio}}, \bibinfo {author} {\bibfnamefont {F.}~\bibnamefont
  {Curciarello}}, \bibinfo {author} {\bibfnamefont {V.~D.}\ \bibnamefont
  {Leo}}, \bibinfo {author} {\bibfnamefont {G.}~\bibnamefont {Fazio}}, \ and\
  \bibinfo {author} {\bibfnamefont {G.}~\bibnamefont {Giardina}},\ }\href@noop
  {} {\bibfield  {journal} {\bibinfo  {journal} {J. Phys. Soc. Jpn.}\ }\textbf
  {\bibinfo {volume} {80}},\ \bibinfo {pages} {094204} (\bibinfo {year}
  {2011})}\BibitemShut {NoStop}%
\bibitem [{\citenamefont {Ajzenberg-Selove}(1988)}]{selov88}%
  \BibitemOpen
  \bibfield  {author} {\bibinfo {author} {\bibfnamefont {F.}~\bibnamefont
  {Ajzenberg-Selove}},\ }\href@noop {} {\bibfield  {journal} {\bibinfo
  {journal} {Nucl. Phys. A}\ }\textbf {\bibinfo {volume} {490}},\ \bibinfo
  {pages} {1} (\bibinfo {year} {1988})}\BibitemShut {NoStop}%
\bibitem [{\citenamefont {Tilley}\ \emph {et~al.}(2002)\citenamefont {Tilley},
  \citenamefont {Cheves}, \citenamefont {Godwin}, \citenamefont {Hale},
  \citenamefont {Hofmann}, \citenamefont {Kelley}, \citenamefont {Sheu},\ and\
  \citenamefont {Weller}}]{till02}%
  \BibitemOpen
  \bibfield  {author} {\bibinfo {author} {\bibfnamefont {D.}~\bibnamefont
  {Tilley}}, \bibinfo {author} {\bibfnamefont {C.}~\bibnamefont {Cheves}},
  \bibinfo {author} {\bibfnamefont {J.}~\bibnamefont {Godwin}}, \bibinfo
  {author} {\bibfnamefont {G.}~\bibnamefont {Hale}}, \bibinfo {author}
  {\bibfnamefont {H.}~\bibnamefont {Hofmann}}, \bibinfo {author} {\bibfnamefont
  {J.}~\bibnamefont {Kelley}}, \bibinfo {author} {\bibfnamefont
  {C.}~\bibnamefont {Sheu}}, \ and\ \bibinfo {author} {\bibfnamefont
  {H.}~\bibnamefont {Weller}},\ }\href@noop {} {\bibfield  {journal} {\bibinfo
  {journal} {Nucl. Phys. A}\ }\textbf {\bibinfo {volume} {708}},\ \bibinfo
  {pages} {3} (\bibinfo {year} {2002})}\BibitemShut {NoStop}%
\bibitem [{\citenamefont {Yamagata}\ \emph {et~al.}(2005)\citenamefont
  {Yamagata} \emph {et~al.}}]{Yamagata05}%
  \BibitemOpen
  \bibfield  {author} {\bibinfo {author} {\bibfnamefont {T.}~\bibnamefont
  {Yamagata}} \emph {et~al.},\ }\href@noop {} {\bibfield  {journal} {\bibinfo
  {journal} {Phys. Rev. C}\ }\textbf {\bibinfo {volume} {71}},\ \bibinfo
  {pages} {064316} (\bibinfo {year} {2005})}\BibitemShut {NoStop}%
\bibitem [{\citenamefont {Brady}\ \emph {et~al.}(1977)\citenamefont {Brady},
  \citenamefont {King}, \citenamefont {Bonner}, \citenamefont {McNaughton},
  \citenamefont {Wang},\ and\ \citenamefont {William}}]{Brady77}%
  \BibitemOpen
  \bibfield  {author} {\bibinfo {author} {\bibfnamefont {F.}~\bibnamefont
  {Brady}}, \bibinfo {author} {\bibfnamefont {N.}~\bibnamefont {King}},
  \bibinfo {author} {\bibfnamefont {B.}~\bibnamefont {Bonner}}, \bibinfo
  {author} {\bibfnamefont {M.}~\bibnamefont {McNaughton}}, \bibinfo {author}
  {\bibfnamefont {J.}~\bibnamefont {Wang}}, \ and\ \bibinfo {author}
  {\bibfnamefont {W.}~\bibnamefont {William}},\ }\href@noop {} {\bibfield
  {journal} {\bibinfo  {journal} {Phys. Rev. C}\ }\textbf {\bibinfo {volume}
  {16}},\ \bibinfo {pages} {31} (\bibinfo {year} {1977})}\BibitemShut {NoStop}%
\bibitem [{\citenamefont {Rae}\ \emph {et~al.}(1984)\citenamefont {Rae},
  \citenamefont {Cole}, \citenamefont {Harvey},\ and\ \citenamefont
  {Stokstad}}]{Rae84}%
  \BibitemOpen
  \bibfield  {author} {\bibinfo {author} {\bibfnamefont {W.~D.~M.}\
  \bibnamefont {Rae}}, \bibinfo {author} {\bibfnamefont {A.~J.}\ \bibnamefont
  {Cole}}, \bibinfo {author} {\bibfnamefont {B.~G.}\ \bibnamefont {Harvey}}, \
  and\ \bibinfo {author} {\bibfnamefont {R.~G.}\ \bibnamefont {Stokstad}},\
  }\href@noop {} {\bibfield  {journal} {\bibinfo  {journal} {Phys. Rev. C}\
  }\textbf {\bibinfo {volume} {30}},\ \bibinfo {pages} {158} (\bibinfo {year}
  {1984})}\BibitemShut {NoStop}%
\bibitem [{\citenamefont {Povoroznyk}(2007)}]{pov07}%
  \BibitemOpen
  \bibfield  {author} {\bibinfo {author} {\bibfnamefont {O.}~\bibnamefont
  {Povoroznyk}},\ }\href@noop {} {\bibfield  {journal} {\bibinfo  {journal}
  {Nuclear Physics and Atomic Energy}\ }\textbf {\bibinfo {volume} {8}},\
  \bibinfo {pages} {131} (\bibinfo {year} {2007})},\ \bibinfo {note} {in
  Ukrainian}\BibitemShut {NoStop}%
\bibitem [{\citenamefont {Tilley}\ \emph {et~al.}(1992)\citenamefont {Tilley},
  \citenamefont {Weller},\ and\ \citenamefont {Hale}}]{till92}%
  \BibitemOpen
  \bibfield  {author} {\bibinfo {author} {\bibfnamefont {D.}~\bibnamefont
  {Tilley}}, \bibinfo {author} {\bibfnamefont {H.}~\bibnamefont {Weller}}, \
  and\ \bibinfo {author} {\bibfnamefont {G.}~\bibnamefont {Hale}},\ }\href@noop
  {} {\bibfield  {journal} {\bibinfo  {journal} {Nucl. Phys. A}\ }\textbf
  {\bibinfo {volume} {541}},\ \bibinfo {pages} {1} (\bibinfo {year}
  {1992})}\BibitemShut {NoStop}%
\bibitem [{\citenamefont {Akimune}\ \emph {et~al.}(2003)\citenamefont {Akimune}
  \emph {et~al.}}]{Akimune03}%
  \BibitemOpen
  \bibfield  {author} {\bibinfo {author} {\bibfnamefont {H.}~\bibnamefont
  {Akimune}} \emph {et~al.},\ }\href@noop {} {\bibfield  {journal} {\bibinfo
  {journal} {Phys. Rev. C}\ }\textbf {\bibinfo {volume} {67}},\ \bibinfo
  {pages} {051302(R)} (\bibinfo {year} {2003})}\BibitemShut {NoStop}%
\end{thebibliography}%

\end{document}